\documentclass[letterpaper, 10 pt, conference]{ieeeconf}  

\usepackage[bottom]{footmisc}

\IEEEoverridecommandlockouts                              
\usepackage[utf8]{inputenc}
\usepackage[T1]{fontenc}
\usepackage{hyperref}
\usepackage{graphicx}
\usepackage{multirow}
\usepackage{graphics} 
\usepackage{epsfig} 
\usepackage{mathptmx} 
\usepackage{mathptmx} 
\usepackage{amsmath} 
\usepackage{amssymb}  

\title{\LARGE \bf
A Deep Knowledge Distillation framework for EEG assisted enhancement of single-lead ECG based sleep staging
}

\author{Vaibhav Joshi$^{2*}$, Sricharan Vijayarangan$^{1,2*}$, Preejith SP$^{1}$, and Mohanasankar Sivaprakasam$^{1,2}$  
\thanks{* Equal Contribution}
\thanks{$^{1}$ are with Healthcare Technology and Innovation Center (HTIC),
        Indian Institute of Technology (IIT-M), India
        {\tt\small sricharanv@htic.iitm.ac.in}}%
\thanks{$^{2}$ is with Department of Electrical Engineering,
        Indian Institute of Technology, Madras (IITM), India
        {}}%
}

\begin{document}
\maketitle

\begin{abstract}


Automatic Sleep Staging study is presently done with the help of Electroencephalogram (EEG) signals. Recently, Deep Learning (DL) based approaches have enabled significant progress in this area, allowing for near-human accuracy in automated sleep staging. However, EEG based sleep staging requires an extensive as well as an expensive clinical setup. Moreover, the requirement of an expert for setup and the added inconvenience to the subject under study renders it unfavourable in a point of care context. Electrocardiogram (ECG), an unobtrusive alternative to EEG, is more suitable, but its performance , unsurprisingly, remains sub-par compared to EEG-based sleep staging. Naturally, it would be helpful to transfer knowledge from EEG to ECG, ultimately enhancing the model's performance on ECG based inputs. Knowledge Distillation (KD) is a renowned concept in DL that looks to transfer knowledge from a better but potentially more cumbersome teacher model to a compact student model.  Building on this concept, we propose a cross-modal KD framework to improve ECG-based sleep staging performance with assistance from features learned through models trained on EEG. Additionally, we also conducted multiple experiments on the individual components of the proposed model to get better insight into the distillation approach. Data of 200 subjects from the Montreal Archive of Sleep Studies (MASS) was utilized for our study. The proposed model showed a 14.3\% and 13.4\% increase in weighted-F1-score in 4-class and 3-class sleep staging, respectively. This demonstrates the viability of KD for performance improvement of single-channel ECG based sleep staging in 4-class(W-L-D-R) and 3-class(W-N-R) classification.

\end{abstract}

\section{INTRODUCTION}

Sleep is a complex dynamic biological process that occurs in multiple cyclical stages. Typically, sleep is studied in sleep medicine by conducting a polysomnography (PSG) study, where multiple bio-signals are acquired during sleep. Electroencephalogram (EEG) signal is the gold standard for sleep studies considering its interpretability with brain activation, the pivot of sleep mechanism. Typically, Sleep staging is done by experts manually for 30/20 s epochs, based on the sleep staging rules and guidelines by AASM (American Academy of Sleep Medicine) \cite{iber2007aasm} or Rechtschaffen and Kales (1968) (the ‘R and K rules’) \cite{rechtschaffen1968manual}. Given that manual sleep staging is time-consuming, several automated sleep staging algorithms, including NN based approaches \cite{perslev2019u}, have been developed in recent history with remarkable performance on par with human accuracy. Conventionally, sleep is classified into five sleep stages:  W, wakefulness; N1, a light sleep period in Non-REM stages; N2, an intermediate stage; N3, a deep sleep stage; and REM Rapid Eye Movement stage. Different stages of sleep are characterized by the different patterns and frequencies observed in the EEG signal during sleep. However, EEG being obtrusive remains impractical in a non-clinical setup. Furthermore, brain-body interaction during sleep implies that other bio-signals can also capture the stages of sleep\cite{abdullah2009correlation}. 

An alternative to the EEG is the Electrocardiogram (ECG) which is not only less obtrusive but also readily integrated into a point of care setup through wearable devices. Q.Li \textit{et al.} \cite{li2018deep} achieved substantial results using ECG signal for sleep staging by extracting ECG Derived Respiration (EDR) and Respiratory Sinus Arrhythmia (RSA) based features for cross spectral spectrogram from a vast PSG dataset (SHHS, CinC, SLPDB) which helped to generalize the model better. In their work, Convolution Neural Network (CNN) was used for feature extraction, which was subsequently fed to  Support Vector Machines (SVM) for sleep stage classification. They achieved an accuracy of 75.4\% and 65.9\% on SLPDB and SHHS, respectively, for 4-class sleep stage classification. Fonseca \textit{et al.} \cite{fonseca2015sleep} used Linear Discriminant (LD) on 80 expert features extracted from ECG and Respiratory Inductance Plethysmography (RIP), which achieved an accuracy of 69\% for 4-class sleep staging and 80\% for 3-class sleep staging. Radha \textit{et al.} \cite{radha2019sleep} used LSTM based temporal model approach for sleep staging on explicitly extracted 132 HRV features from ECG signal and achieved 77\% accuracy. N. Sridhar \textit{et al.} \cite{sridhar2020deep} also achieved 77\% accuracy with 2 stage CNNs by ECG-derived heart rate on a vast dataset comprising SHHS, MESA and CinC. Despite the increasing body of work, studies show that the performance of ECG based sleep staging remains vastly inferior to EEG based sleep staging \cite{perslev2019u}\cite{sun2020sleep}. It would be immensely beneficial to combine the benefits of unobtrusive monitoring with improved accuracy. While multi-modal fusion methods for sleep staging get features of both signals and have been shown to provide improved accuracy \cite{fonseca2015sleep}\cite{sun2020sleep}, this setup requires multiple signal acquisition during inference.  Thus, it is of great interest to employ the primary modality, EEG, to inform ECG based modelling without significant overhead during inference.

Knowledge Distillation (KD) in deep networks has recently garnered much traction to efficiently compress and transfer relevant information from extensive networks to more compact ones. Hinton \textit{et al.} \cite{hinton2015distilling} proposed a response based KD method to distil and transfer information from the softmax layer of a larger size teacher model to a student model to achieve better generalization. Romero \textit{et al.} \cite{romero2014fitnets} proposed a feature-based KD by deriving knowledge from the intermediate feature maps rather than the softmax layers. Zagoruyko \textit{et al.} \cite{komodakis2017paying} introduced the concept of transferring attention maps of the intermediate feature layers to enhance the distillation process. This was also extended across modalities spawning a separate subfield of cross-modal distillation \cite{gou2021knowledge}.
Deriving inspiration from these approaches, we propose a cross-modal KD approach combining response based and feature-based distillation that would enable multi-modal training of EEG and ECG while allowing for unimodal testing, enabling the usage of ECG alone during inference. To the best of our knowledge, this is the first study designed to validate the viability of KD to enhance ECG based sleep staging. Additionally, we compare the proposed model against individual components of the KD framework and subsequently demonstrate its efficacy.

\section{METHODS}
\subsection{Problem Formulation}


Let $x_{eeg}$ $ \in\ \mathbb{R}^{\tau S} $\ and $x_{ecg}$ $ \in\ \mathbb{R}^{ \tau S} $\ be the EEG and ECG waveforms, sampled at rate \textit{S} for $ \tau $ seconds, respectively. Let $F(eeg; \theta_{eeg})$ and $F(ecg; \theta_{ecg})$ be the models which take in T fixed-length connected
segments, each of length \textit{i}, from $x_{eeg}$ and $x_{ecg}$ respectively. Let \textit{e}  be the frequency at which the signal is segmented,where the objective is to independently map $x_{eeg}$ and $x_{ecg}$ to [\textit{e} * $\tau$] labels,
where each label is based on \textit{i= S/\textit{e}} sampled points. While thirty second intervals are usually considered (i.e., \textit{e} = 1/30 Hz), the architecture we adapt is capable of handling different frequencies during inference. Specifically, the model $F(eeg; \theta_{eeg})$ and $F(ecg; \theta_{ecg})$
maps $x_{eeg}$ and $x_{ecg}$ to class confidence scores for predicting K
classes in all T segments. The loss function optimized by the teacher model, $F(eeg; \theta_{eeg})$, is the Weighted Cross Entropy(WCE) defined as:

\begin{equation} \label{eq:1}
L(x_{eeg}, y) = \sum_{i=1}^{T} 1/\sum_{i=1}^{T} (-w_{yi})  * l_{eeg}^i
\end{equation}

where $w_{yi}$ is the class weight based on the number of samples belonging to a particular class in the training set and

\begin{equation} \label{eq:2}
l_{eeg}^i = w_{y_i} * log(exp(x_{eeg}^{(i,y_i)}) / \sum_{c=1}^{K}  exp(x_{eeg}^{(i,c)})
\end{equation}

The next process involves feature based distillation of the intermediate attention maps from trained teacher model, $F(eeg; \theta_{eeg})$, to the untrained student model, $F(ecg; \theta_{ecg})$. To obtain effective information distillation, we adopt the following Attention Transfer (AT) loss \cite{komodakis2017paying}:

\begin{equation} \label{eq:3}
    L_{AT} = \sum_{j \in I}||\frac{Q_{ecg}^{j}}{||Q_{ecg}^{j}||_{2}} - \frac{Q_{eeg}^{j}}{||Q_{eeg}^{j}||_{2}}||_{2}
\end{equation}

where $Q_{ecg}^{j}$=$vec(F_{ecg}(A_{ecg}^{j}))$ and $Q_{eeg}^{j}$=$vec(F_{eeg}(A_{eeg}^{j}))$ represent the $j$-th pair of student and teacher attention maps in a vectorized form, $I$ denote the set of teacher-student convolution layers which is selected for AT. In our framework, $j$ iterates through the feature map across the whole architecture, distilling the attention maps from all the layers. 

Subsequently, the pretrained student model $F(ecg; \theta_{ecgpre})$, optimizes the sum of WCE and the response based distillation loss defined as:

\begin{equation} \label{eq:4}
L(x_{ecg}, y) = (1-\alpha)*\sum_{i=1}^{T} 1/\sum_{i=1}^{T} (-w_{yi})  * l_{ecg}^i +  \alpha T_d^2 * l_{dist}^i
\end{equation}

where alpha is the weight between distillation loss and classification loss, $T_{d}$ is the softmax temperature and   
\begin{equation} \label{eq:5}
l_{ecg}^i = w_{y_i} * log(exp(x_{ecg}^{(i,y_i)}) / \sum_{c=1}^{K}  exp(x_{ecg}^{(i,c)})
\end{equation}

\begin{equation} \label{eq:6}
l_{dist}^i = KLDiv(p(x_{ecg}, T_d), p(x_{eeg}, T_d))
\vspace{-1.35em}
\end{equation}

\begin{equation} \label{eq:7}
p(x,T_d) = log(exp(x^{(i,y_i)}/T_d) / \sum_{c=1}^{K}  exp(x^{(i,c)}/T_d)
\end{equation}

Where $KLDiv$ is the Kulback Leibler Divergence.

\begin{figure}[t]
    \centering
      \framebox{\parbox{3in}{
      \centering
      \includegraphics[width=0.4\textwidth]{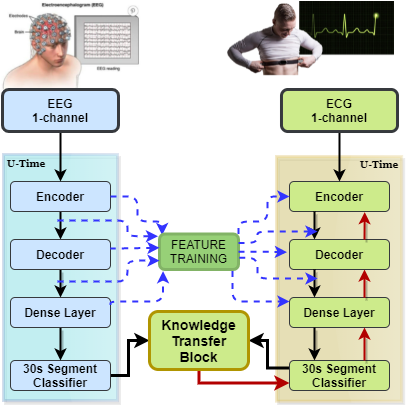}
}}
    
    \caption{Proposed Knowledge Distillation framework}
    \label{fig:1}
\vspace{-1.2em}
\end{figure}

\subsection{Architectural Details}

The base architecture for both the teacher, $F(eeg; \theta_{eeg})$ and the student, $F(ecg; \theta_{ecg})$ is partially adapted from U-Time \cite{perslev2019u}, considering three major reasons:
\begin{enumerate}
    \item  \textbf{Optimized on EEG} - U-Time was originally optimized for EEG based sleep staging, thereby maximizing the feature transfer from EEG to ECG.
    \item  \textbf{Segmentation at inference time} - A U-Time model trained to segment for every 30s may be used to output sleep stages at a higher frequency, i.e., every 20s, at inference time.
   \item \textbf{Benefit of being fully convolutional} - U-Time, as opposed to other architectures, can be applied across multiple datasets which contain variations without any architecture or hyperparameter tuning.
\end{enumerate}

Minor changes to the architecture were done to optimize the ECG baseline, which must be improved effectively. An additional layer increased the depth of the network in the Encoder and Decoder of the original network. Furthermore, differences in sampling rates implying different window sizes had required the modification of the spatial reduction occurring at each max-pooling layer. 

\subsection{Dataset Description}

In this study, we utilize the Montreal Archive of Sleep Studies (MASS) \cite{o2014montreal} dataset comprising of whole‐night recordings from 200 participants [97 males (aged 42.9 ± 19.8 years) and 103 females (aged 38.3 ± 18.9 years); age range: 18–76 years]; organized according to their research and acquisition protocols into five subsets of recordings, SS1-SS5. All subjects included in this cohort are healthy controls, except for 15 included in SS1, diagnosed with Mild Cognitive Impairment (MCI) according to a neuropsychological evaluation. We used EEG and ECG data from all the subjects, i.e. 200 subjects in total. From the EEG electrodes(positioned according to the international 10-20 system), C3-A2/C4-A1 EEG electrodes and Lead 1 ECG were used for our study. The data was resampled down to 200 Hz from their respective original sampling rate to maintain uniformity. We also converted the data with 20s segment annotations to 30s segment by including 5-second data on both extremes. We modified sleep stage annotations N1, N2 into Light Sleep(L) and N3, N4 into Deep Sleep(D) for our 4-class (W-L-D-R) classification and combined N1, N2, N3, N4 into NREM(N) for our 3-class(W-N-R) classification.
 
\subsection{Experimental Procedures}
The 200 subject data was split subject-wise into train-validation-test sets with 80:10:10 split for W-R-N and W-R-L-D staging, which ensured that no data from the same subject existed across the splits. For all the experiments, the best model was saved based on the validation metrics inspected during training and finally evaluated on the holdout test-set after completion of training. The metrics used to evaluate the model on the holdout test set were weighted F1 score and accuracy as done in \cite{perslev2019u}. The weighted-F1 score calculated the metric for each class separately and averaged the metrics across classes, weighting each class by its support (tp + fn) to counter the inherent imbalance in the sleep stages. We identically conducted all the experiments for both W-L-D-R and W-N-R classification following the framework shown in Fig.\ref{fig:1}. Baselines were trained by optimizing on WCE loss as in $Eq.\ref{eq:1}$ and the KD experiment optimizing the loss in $Eq.\ref{eq:4}$. Additionally, we conducted two experiments to understand the individual components of our proposed model. The primary components of all the experiments involve two major steps:
\begin{enumerate}
    \item Feature Training (Step 1): $F(ecg; \theta_{ecg})$ optimizes the loss ($Eq.\ref{eq:3}$) between EEG and ECG features with EEG weights being frozen. This trains the feature maps in the ECG model to mimic the feature maps of the EEG model.
    \item Final Training (Step 2): $F(ecg;\theta_{ecgpre})$ from Step 1 is further trained on distillation loss($Eq.\ref{eq:4}$) for optimizing the weights of the ECG. The value of T was chosen as 1, as given in \cite{hinton2015distilling} and alpha was empirically chosen to provide ideal weights to distillation and classification loss.
\end{enumerate}
The mechanism followed of the distillation methods is as follows:

\begin{enumerate}
    \item \textbf{AT+SD+CL(proposed method)}: Step 1 is executed followed by step 2 with $ \alpha = 0.5$ in $Eq.\ref{eq:4}$.
    \item \textbf{AT+CL(ablation method)}: Step 1 is executed followed by step 2 with $ \alpha = 0$ in $Eq.\ref{eq:4}$, which results in training on the classification loss independently in $Eq.\ref{eq:1}$.
    \item \textbf{SD+CL(ablation method)}: Only step 2 is executed with $ \alpha = 0.5$ in $Eq.\ref{eq:4}$.
\end{enumerate}
\footnote{AT: Attention Transfer; SD: Softmax Distillation; CL: Classification Loss}
Each configuration was trained for 150 epochs with learning  rate(LR) of $10^{-3}$. The model was implemented in Pytorch on an Nvidia GTX3090Ti 24GB GPU. 
\footnote{Code will be provided on acceptance}



\section{RESULTS}
\begin{table}[b]
\vspace{-1.5em}
\caption{Performance of KD and its components
\label{Result_Table}}
\centering
\begin{tabular}{|c|lcc|}
\hline
\textbf{No. of classes} & \multicolumn{1}{c|}{\textbf{Experiments}} & \multicolumn{1}{c|}{\textbf{weighted F1}} & \textbf{Accuracy} \\ \hline
\multirow{5}{*}{W-R-L-D} & EEG Baseline & 0.85          & 0.85          \\
                         & ECG Baseline & 0.45          & 0.44          \\
                         & SD + CL      & \textbf{0.51} & \textbf{0.51} \\
                         & AT + CL      & 0.50          & 0.50          \\
                         & AT + SD + CL & 0.50          & 0.49          \\ \hline
\multirow{5}{*}{W-R-N}   & EEG Baseline & 0.90          & 0.90          \\
                         & ECG Baseline & 0.58          & 0.56          \\
                         & SD + CL      & 0.61          & 0.60          \\
                         & AT + CL      & \textbf{0.66} & \textbf{0.66} \\
                         & AT + SD + CL & 0.64          & 0.63          \\ \hline
\end{tabular}
\vspace{-1.5em}
\end{table}

Given that the primary intention of this work is to establish the efficacy of KD, we compare the distillation frameworks against their respective baseline. Table.\ref{Result_Table} shows the performance of our models on the holdout test data. Increments in performance were observed for both weighted-F1-score and accuracy metrics for the proposed AT+SD+CL distillation method as well as ablation methods for both 3-class and 4 class sleep staging. The SD+CL model was the best performing model for 4-class, where the weighted-F1 score improved to \textbf{0.51} from \textbf{0.45} of ECG baseline (weighted-F1 improved by \textbf{14.3\%}, Accuracy improved by 15.6\%). For 3-class, AT+CL was the best performing model with weighted-F1 improved to \textbf{0.66} from \textbf{0.58} of ECG baseline( weighted-F1 improved by \textbf{13.4\%}, Accuracy improved by 18.1\%). However, all the distillation models outperformed the ECG baseline model, which supports our idea of applying KD for ECG based sleep staging. 

\begin{table}[t]
\caption{KD methods class wise Results} \label{Classwise_Table}
\centering
\begin{tabular}{|l|cccc|ccc|}
\hline
\multicolumn{1}{|c|}{\multirow{2}{*}{}} &
  \multicolumn{4}{c|}{\textbf{4 class F1 score}} &
  \multicolumn{3}{c|}{\textbf{3 class F1 score}} \\ \cline{2-8} 
\multicolumn{1}{|c|}{} &
  \multicolumn{1}{c|}{\textbf{W}} &
  \multicolumn{1}{c|}{\textbf{L}} &
  \multicolumn{1}{c|}{\textbf{D}} &
  \textbf{R} &
  \multicolumn{1}{c|}{\textbf{W}} &
  \multicolumn{1}{c|}{\textbf{N}} &
  \textbf{R} \\ \hline
\textbf{EEG Baseline} & 0.89 & 0.86          & 0.81 & 0.82 & 0.89 & 0.93          & 0.80 \\
\textbf{ECG Baseline} & 0.57 & 0.47          & 0.30 & 0.40 & 0.51 & 0.65          & 0.40 \\
\textbf{SD + CL}      & 0.54 & \textbf{0.61} & 0.31 & 0.35 & 0.53 & 0.70          & 0.34 \\
\textbf{AT + CL}      & 0.54 & 0.57          & 0.34 & 0.40 & 0.52 & \textbf{0.77} & 0.37 \\
\textbf{AT + SD + CL} & 0.57 & 0.56          & 0.29 & 0.41 & 0.52 & 0.73          & 0.39 \\ \hline
\end{tabular}
\vspace{-1.9em}
\end{table}
Table.\ref{Classwise_Table} gives insight into the class-wise performances of distillation methods. The best improvement was observed for Light sleep(L) in 4-class staging where weighted-F1 improved from \textbf{0.47} to \textbf{0.61} by SD+CL ablation method. For 3-class staging, the best improvement was observed for NREM class with weighted-F1 increasing from \textbf{0.65} to \textbf{0.77} by the AT+CL ablation method. Noticeably, other classes have marginally underperformed. This is potentially due to the imprecise feature training owing to class imbalance during training. Additionally, the proposed AT+SD+CL distillation showed reasonable improvement across all classes in both 4-class and 3-class, thus relatively robust against the class imbalance during feature learning.
\vspace{-0.5em}
\section{DISCUSSION}
From the above-indicated results, we demonstrate the viability of the KD approach for ECG based sleep staging as the proposed model outperformed the ECG baseline significantly. However, combining response-based (SD) and feature-based (AT) distillation does not always yield improved results over using these distillation modes separately. This highlights the complexity of the interplay between these two modes of KD and suggests that combining these two distillation modes may require independent optimization.

\begin{figure}[t]
    \centering
      \framebox{\parbox{3in}{
      \centering
      \includegraphics[width=0.4\textwidth]{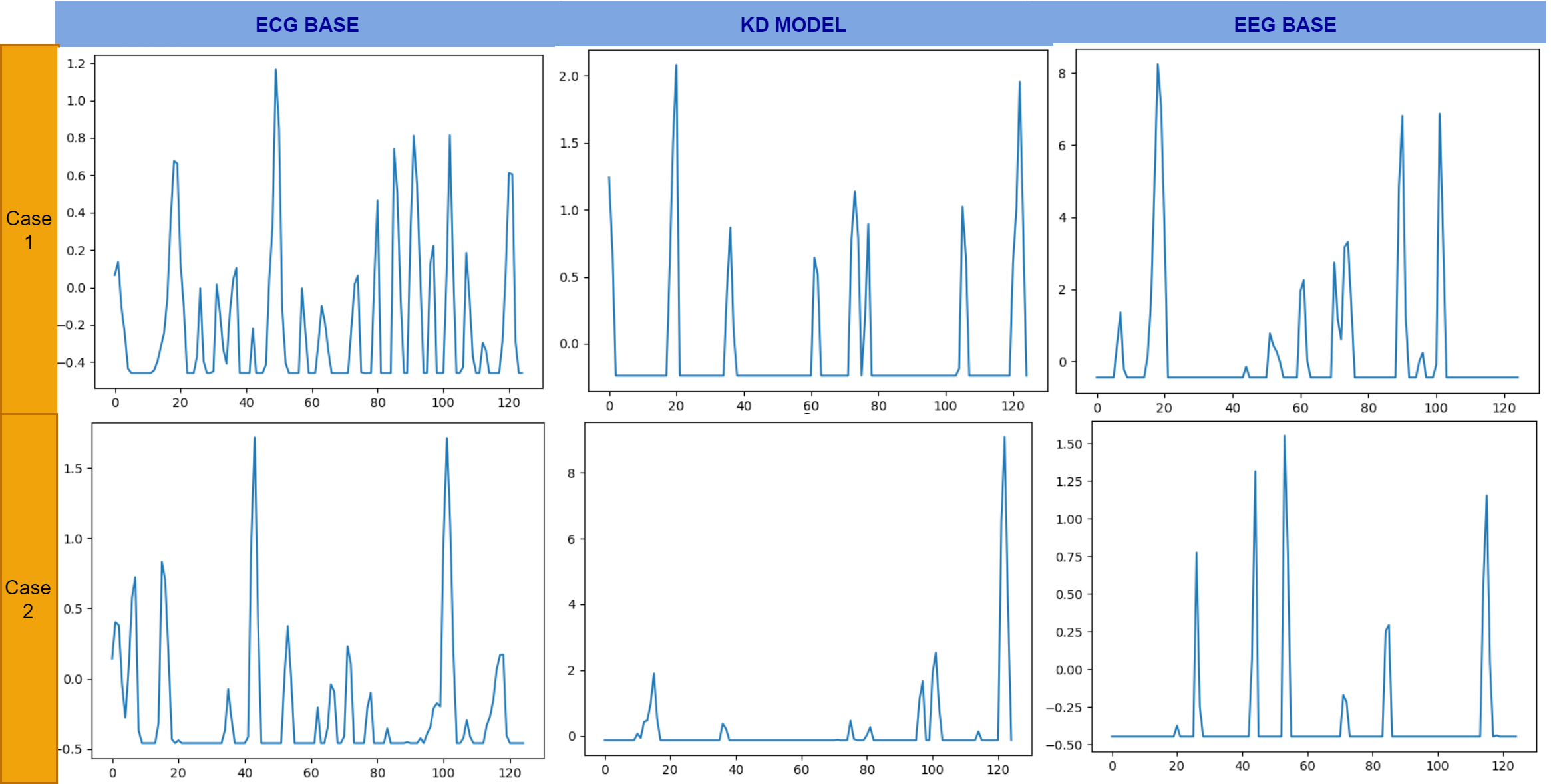}
}}
    
    \caption{Bottleneck layer feature for case1:KD model correct,ECG base incorrect; case2:ECG baseline correct, KD model incorrect.}
    \label{fig:Feature_plots}
\vspace{-1.5em}
\end{figure}

Figure.\ref{fig:Feature_plots} aids in analyzing the feature learning in the bottleneck layer, which contains the most compressed representation. The figure shows two scenarios for 4-class sleep staging; 
case 1, when the proposed model predicts accurately, but ECG baseline predicts incorrectly;
case 2, when the KD model predicts incorrectly, but ECG baseline predicts accurately.
It is evident that distilled model's features in case 1 optimal feature learning, ultimately improving performance. However, in case 2, the distilled model's feature distinctively differs from the ECG and the EEG baseline, resulting in misguided feature learning. This could be partially attributed to the class imbalance which may lead to asymmetric distillation as can be seen in Table \ref{Classwise_Table}.

Despite the promising performance improvement brought about by KD, there are few limitations in our study. Firstly, The baseline ECG model utilized in this paper can be improved by using temporal models like Long Short Term Memory networks (LSTM) which capture sparsely distributed features over time.  Furthermore, using additional unobtrusive modalities along with ECG have been shown to enhance sleep staging. Previous works \cite{sridhar2020deep}\cite{li2018deep} which achieved notable performance on ECG based sleep staging have been trained on an extensively large dataset (>4000 records), whereas our study used a relatively compact dataset. Future work would involve translating the benefits of KD to multiple DL architectures, ultimately improving overall accuracy.

\section{CONCLUSIONS}

We conclude this study with the following contributions which we believe broadens the current knowledge in ECG based sleep staging.
    We validate the potency of a KD framework for performance improvement of ECG in sleep staging.
    We analyze the individual components of the KD framework and subsequently dissect the features in the bottleneck layer to understand the reasons for improved performance.   

\addtolength{\textheight}{-12cm}   


\bibliographystyle{IEEEtran}
\bibliography{ref}
\end{document}